\documentclass[%
 aip,
cp,  % Conference Proceedings
 amsmath,amssymb,%nobibnotes,
 reprint,%
]{revtex4-2}

\usepackage{graphicx}% Include figure files
\usepackage{dcolumn}% Align table columns on decimal point
\usepackage{bm}% bold math
\usepackage[utf8]{inputenc}
\usepackage[T1]{fontenc}
\usepackage{mathptmx} 

\begin{document}
\sloppy

\title{Neutron Yield Calculation from ($\alpha,n$) Reactions with SOURCES4}% Force line breaks with \\

\author{Vitaly A. Kudryavtsev} % Write as First name Surname
 \email[Corresponding author: ]{v.kudryavtsev@sheffield.ac.uk}
\author{Piotr Krawczun}%
% \email{second.author@institution.edu.}
\author{Rayna Bocheva}%
\affiliation{Department of Physics and Astronomy, University of Sheffield, Sheffield, S3 7RH, UK
% Force line breaks with \\ if necessary
}

%\author{Another's Name}
 %\email{third.author@anotherinstitution.edu}
%\affiliation{%
%Second institution and/or address% Force line breaks with \\ if necessary
%
%\affiliation{You would list an author's second affiliation (if applicable) here.}

\date{\today} % It is always \today, today, but any date may be explicitly specified
              % Not printed for conference proceedings

\begin{abstract}
Neutrons can induce background events in underground experiments looking for rare processes. Neutrons in a MeV range in deep underground laboratories are produced in spontaneous fission processes and ($\alpha,n$) reactions. A number of computer codes are available to calculate cross-sections of ($\alpha,n$) reactions, branching ratios to various states and neutron yields. We have used the SOURCES4 code to calculate neutron yields and energy spectra with input cross-sections and branching ratios taken from experimental data and calculations based on EMPIRE2.19/3.2.3 and TALYS1.9. We report here a comparison of SOURCES4 calculations with experimental data obtained with alpha beams with fixed energies.
\end{abstract}

\maketitle

\section{\label{sec:intro}Introduction}

Particles produced in radioactive processes may affect sensitivity of low-background experiments, usually constructed in underground laboratories and looking  for rare events such as dark matter or neutrino interactions.  A particular problem is caused by neutrons produced in spontaneous fission and ($\alpha,n$) reactions that can mimic, for instance, WIMP-induced events in direct dark matter searches if a neutron scatters only once in the detector volume and does not leave any signal in veto systems. 

Several computer codes exist to calculate neutron yields and energy spectra form these processes (see, for instance, Refs.~\cite{usd,neucbot,mendoza2019,sources4}). Spontaneous fission (SF) is well described by the parameterisation suggested by Watt~\cite{watt} with parameters tuned to the measurements. The neutron yield from SF does not depend on the material where the neutron emission happens, but only on the concentration of  $^{238}$U (other radioactive isotopes give negligible neutron yield from this process). The main problem is then in accurate calculation neutron background caused by ($\alpha,n$) reactions where alpha particles are produced in the decay chains of radioactive isotopes of uranium and thorium. The codes usually use as inputs cross-sections of these reactions and transition probabilities to excited states together with energy losses of alphas as they travel through the material until they stop. Another option is to take neutron spectra directly from the libraries such as JENDL~\cite{jendl} or TENDL~\cite{talys}.

A comparison between different codes to calculate neutron yields and spectra and experimental data has been given in several papers, see for instance Refs.~\cite{neucbot,mendoza2019,sources4,fernandes,scorza}. In this paper, we report the new calculations of neutron production with the SOURCES4 code that includes `optimised' cross-sections and transition probabilities. This `optimisation' includes a combination of recent experimental data for the cross-sections and a model where the data are not available. The same model is also used in the calculation of branching ratios to different states (transition probabilities). We focus here on ($\alpha,n$) reactions caused by alphas up to 10 MeV. These energies are typical for alphas from the radioactive decay chains of $^{238}$U, $^{235}$U and $^{232}$Th which are the main contributors to neutron background in underground experiments. 

In the next Section we briefly describe the SOURCES4 code and the cross-sections used. Then we show the comparison of our calculations with experimental data and other codes.

\section{The SOURCES4 code to calculate neutron production ($\alpha,n$) reactions}
\label{sec:sources4a}

The nuclear physics code SOURCES4~\cite{sources4} has been used for a long time in a number of applications.  
The code libraries contain alpha emission lines from radioactive isotopes, cross-sections of ($\alpha,n$) reactions from calculations and experimental data, probabilities of transitions to different final states of the daughter nucleus and parameterisations for energy losses of alphas in different materials. The code calculates the neutron production rates (or yields) and energy spectra of emitted neutrons for several types of problems. We will consider here the thick target neutron yield from radioactive isotopes in a homogeneous material where the size of the material sample is much bigger than the range of alphas so edge effects can be neglected, and the neutron production by monoenergetic alphas in a beam where, again, the size of a material sample is much bigger than the range of alphas. The most recent version of the code is SOURCES4C~\cite{sources4} but for historical reasons we are using the older version SOURCES4A \cite{sources4a}. Previous tests and release notes confirmed that both versions give the same results if the same cross-sections and transition probabilities to excited states are used in both versions. A big advantage of the code is in its flexibility so that a user can choose what cross-section of ($\alpha,n$) reaction is used for a particular isotope in a material sample. The user can also add more cross-sections to the library. 

The original code calculates neutron production for alpha energies up to 6.5 MeV and is not fully suitable for calculation of neutrons from radioactive processes that involve alphas with energies up to about 9 MeV. The original code SOURCES4A was modified to remove the 6.5 MeV energy cut and an updated version now allows calculations of neutron production from alphas with energies up to 10~MeV~\cite{carson,tomasello2008}. Libraries of cross-sections and branching ratios were updated with calculated values from the EMPIRE2.19 code~\cite{empire} and extended to alpha energies up to 10~MeV~\cite{carson,tomasello2008,lemrani,tomasello2010,tomasello-thesis}. Cross-sections and branching ratios were calculated with a 0.1~MeV step in alpha energy. All these changes have been made to the version SOURCES4A and, since no changes in results were observed with the newer version of the code, SOURCES4A is continued to be used in most calculations. A comparison of cross-sections from EMPIRE2.19 with some experimental data was published in Refs.~\cite{tomasello2008,tomasello-thesis} and the results from the modified SOURCES4A code were used in a number of dark matter experiments (see, for example, Refs.~\cite{eureca,edelweiss,lz,xenon1t}).

The user input to SOURCES4A includes either the energy of an alpha particle or the $Z$ and $A$ of the radioactive isotope (or several isotopes in the case of decay chains, for instance) with the number of atoms per unit volume. The user should also specify material composition (where the alpha sources are located) and isotopic composition for each element (only isotopes with cross-sections present in the code library can be included). 

The output of SOURCES4A includes several files that return the neutron yield and spectra for the sum of the ground and all excited states, as well as neutron spectra for individual states. The output also includes neutron production spectra on each isotope in the material sample. In the case of decay chains, neutron production from individual radioactive isotopes on each isotope in the material sample is also returned by the code. The code does not generate gammas produced from de-excitation of a nucleus in the final state but neutron spectra are calculated for individual states. 

Recently, the libraries of SOURCES4A have been updated to include some cross-sections calculated with TALYS1.9~\cite{talys} and the newer version of EMPIRE3.2.3~\cite{empire3.2}. The comparison of cross-sections from TALYS1.9, EMPIRE2.19, EMPIRE3.2.3 and experimental data has been presented in Ref.~\cite{vk2020}. Ref.~\cite{vk2020} also includes comparison of neutron yields calculated using SOURCES4A with different cross-sections based on calculations using TALYS1.9, EMPIRE2.19/3.2.3 and some experimental data. 

Based on the results reported in Ref.~\cite{vk2020}, the libraries of cross-sections and branching ratios in SOURCES4A have been updated so the most reliable data are used when available; when the data are scarce or do not exist, the model calculations from either TALYS1.9 or EMPIRE2.19/3.2.3 are used for these isotopes and/or alpha energies. The same model is used for branching ratios since there are very few data on this in literature. Sometimes this approach (a combination of most reliable data and model calculations) is called `evaluated' libraries. We will refer to this approach and cross-sections as `optimised'.

\section{SOURCES4 results and comparison with other codes and experimental data}
\label{sec:results}

Here we present the results from SOURCES4A calculations using 'optimised' cross-sections and branching ratios. Where possible the measured cross-sections have been used in the calculations of neutron yields. If different sets of data are not in agreement, or there are no data for a particular isotope, the model from either TALYS1.9 or EMPIRE2.19/3.2.3 is used. Branching ratios were calculated using TALYS1.9 or EMPIRE2.19/3.2.3. The results of these calculations for a number of elements are compared to different data sets that were obtained from $\alpha$-particle beams interacting with thick targets composed of different elements/isotopes. This comparison provides another independent test of neutron yield calculations. 

\begin{figure}[ht!]
\begin{minipage}{7cm}
\includegraphics[width=7cm]{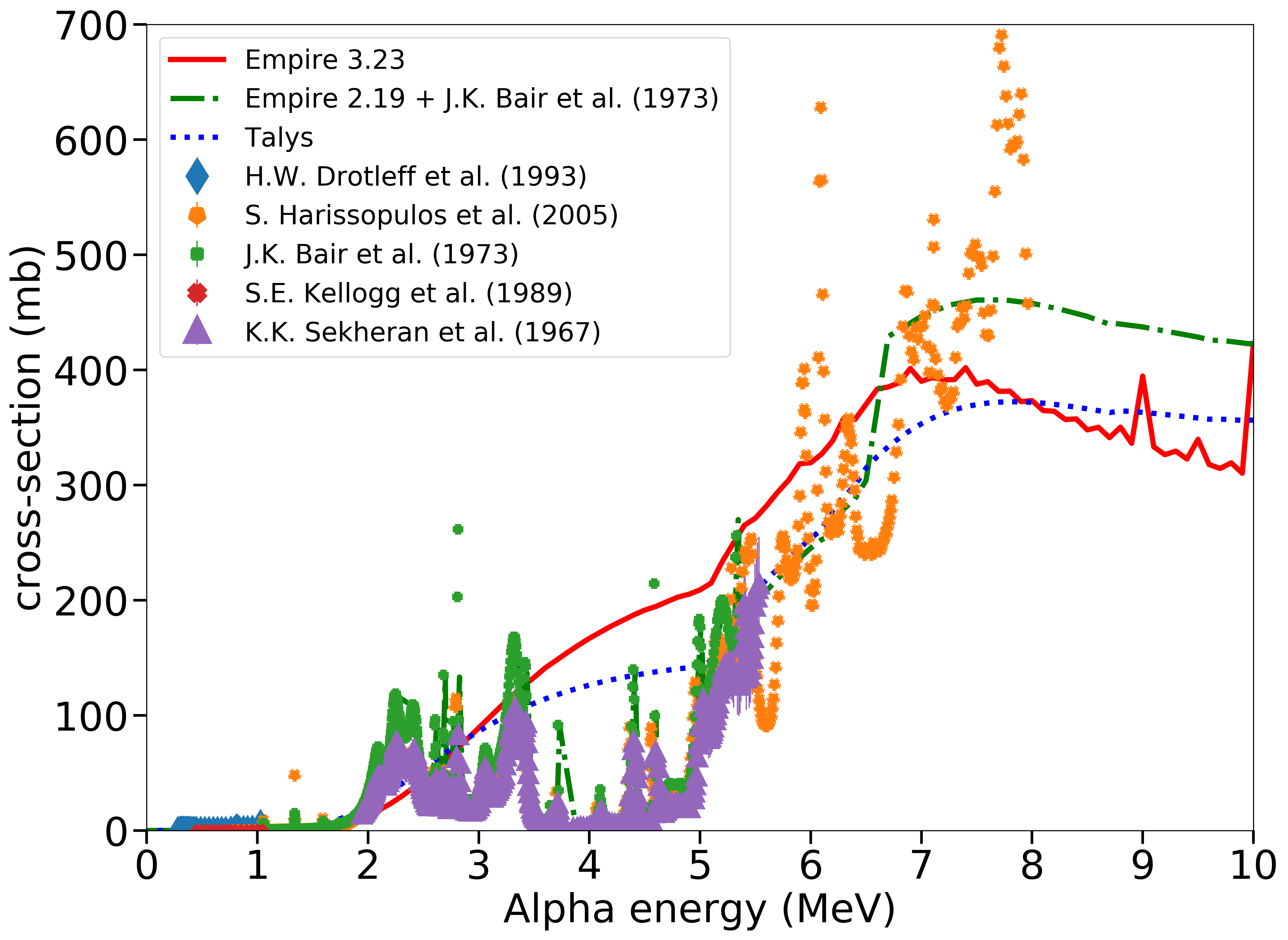}
\caption{\label{fig:cs-C}
Cross-section of an ($\alpha,n$) reaction on $^{13}$C~\cite{zakhary}. Experimental data from Refs.~\cite{drotleff1993,harissopulos2005,bair1973,kellogg1989,shekharan1967} are shown together with EMPIRE2.19/3.2.3 and TALYS1.9 models. Cross-section marked as `EMPIRE2.19 + J.K. Bair et al. (1973)' includes measurements from Ref.~\cite{bair1973} and calculations with EMPIRE2.19 above 5.5~MeV.}
\end{minipage}\hspace{0.5cm}
\begin{minipage}{8.5cm}
\includegraphics[width=8.5cm]{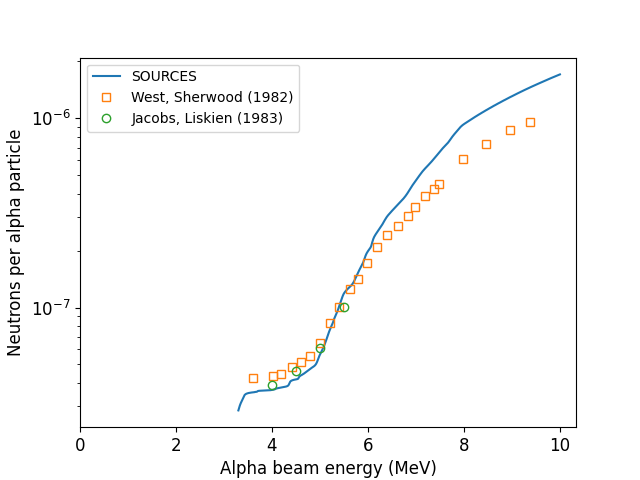}
\caption{\label{fig:yield-C}
Neutron yield as a function of alpha energy for natural carbon. `Optimised' cross-sections are used in SOURCES4A: experimental data from Ref.~\cite{harissopulos2005} up to 8 MeV and the model from TALYS1.9~\cite{talys} above 8~MeV and for branching ratios. The measurements for natural carbon have been taken from Ref.~\cite{west1982}.}
\end{minipage}
\end{figure}

The cross-section for ($\alpha,n$) reaction on $^{13}$C is plotted in Figure~\ref{fig:cs-C}~\cite{zakhary}. The threshold for ($\alpha,n$) reactions on $^{12}$C is above energies of $\alpha$ lines from radioactive decay chains of uranium and thorium so only $^{13}$C yield contributes to neutron production in carbon. Data were taken from Refs.~\cite{drotleff1993,harissopulos2005,bair1973,kellogg1989,shekharan1967}. Calculations with TALYS1.9 and EMPIRE3.2.3 are also shown. The cross-section marked as `EMPIRE2.19 + J.K. Bair et al.\ (1973)' includes measurements from Ref.~\cite{bair1973} and calculations with EMPIRE2.19 above 5.5~MeV.

Figure~\ref{fig:yield-C} shows the neutron yield from carbon as a function of $\alpha$-particle energy. The measurements for natural carbon have been reported in Ref.~\cite{west1982}. SOURCES4A calculations use the cross-sections for $^{13}$C from Ref.~\cite{harissopulos2005} up to 8 MeV. Above this energy and for branching ratios the model from TALYS1.9~\cite{talys} was implemented (see also Ref.~\cite{vk-lrt2022} for the comparison of data and models for $^{13}$C and some other isotopes). Although the cross-section for ($\alpha,n$) reaction on $^{13}$C as input to SOURCES4A has been taken from the measurements, the resulting neutron yield from calculation and its trend with energy are not in good agreement with data, in particular above 6 MeV. We note, however, that statistical models and evaluated libraries~\cite{talys,empire,empire3.2,jendl} do not predicted so high rise of cross-sections above 6~MeV as in Ref.~\cite{harissopulos2005} so calculated neutron yield may indeed be lower and agree better with the alpha beam measurements.

\begin{figure}[ht!]
\begin{minipage}{7.5cm}
\includegraphics[width=7.5cm]{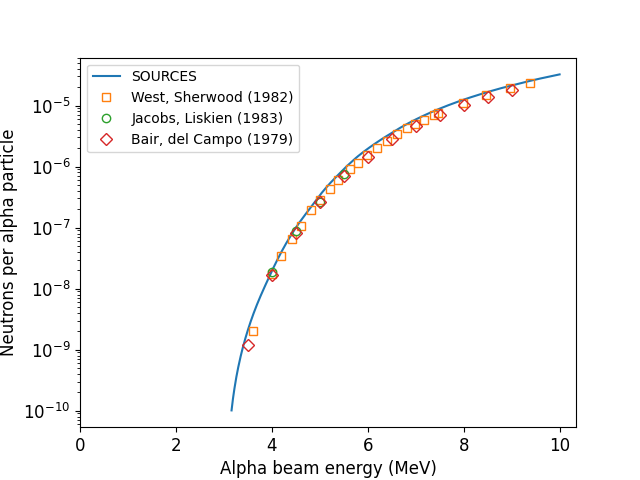}
\caption{\label{fig:nyield-Al}
Neutron yield as a function of alpha energy for aluminium. The cross-sections as input to SOURCES4A were calculated with TALYS1.9. Experimental data were from Ref.~\cite{west1982}}.
\end{minipage}\hspace{0.5cm}
\begin{minipage}{7.5cm}
\includegraphics[width=7.5cm]{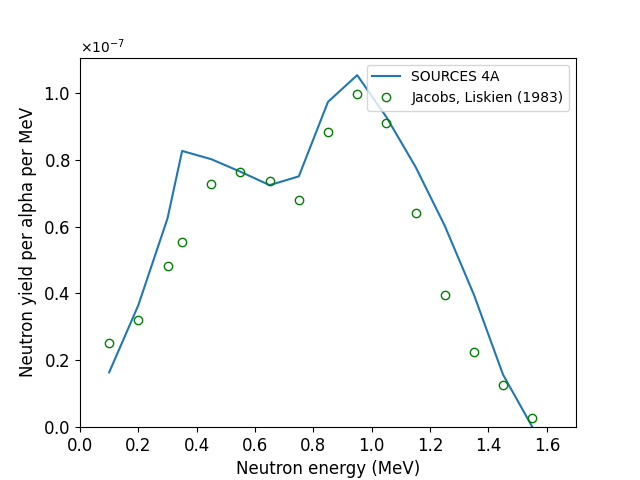}
\caption{\label{fig:nspectra-Al}
Neutron energy spectrum from 4.5~MeV alphas stopping in aluminium. The cross-sections as input to SOURCES4A were calculated with TALYS1.9. Experimental data were from Ref.~\cite{jacobs}.}
\end{minipage}
\end{figure}

Neutron yield as a function of alpha energy from aluminium is shown in Figure~\ref{fig:nyield-Al}. The cross-section from the TALYS1.9 model was used as input to SOURCES4A due to inconsistency between available data sets and lack of the data above 6 MeV (see the discussion in Ref.~\cite{vk-lrt2022}). The resulting neutron yield from calculations is in good agreement with data. Neutron energy spectra from SOURCES4A calculations for 4.5~MeV alpha beam also agree quite well with experimental data (see Figure~\ref{fig:nspectra-Al}).

\begin{figure}[ht!]
%\begin{minipage}{8cm}
\includegraphics[width=15cm]{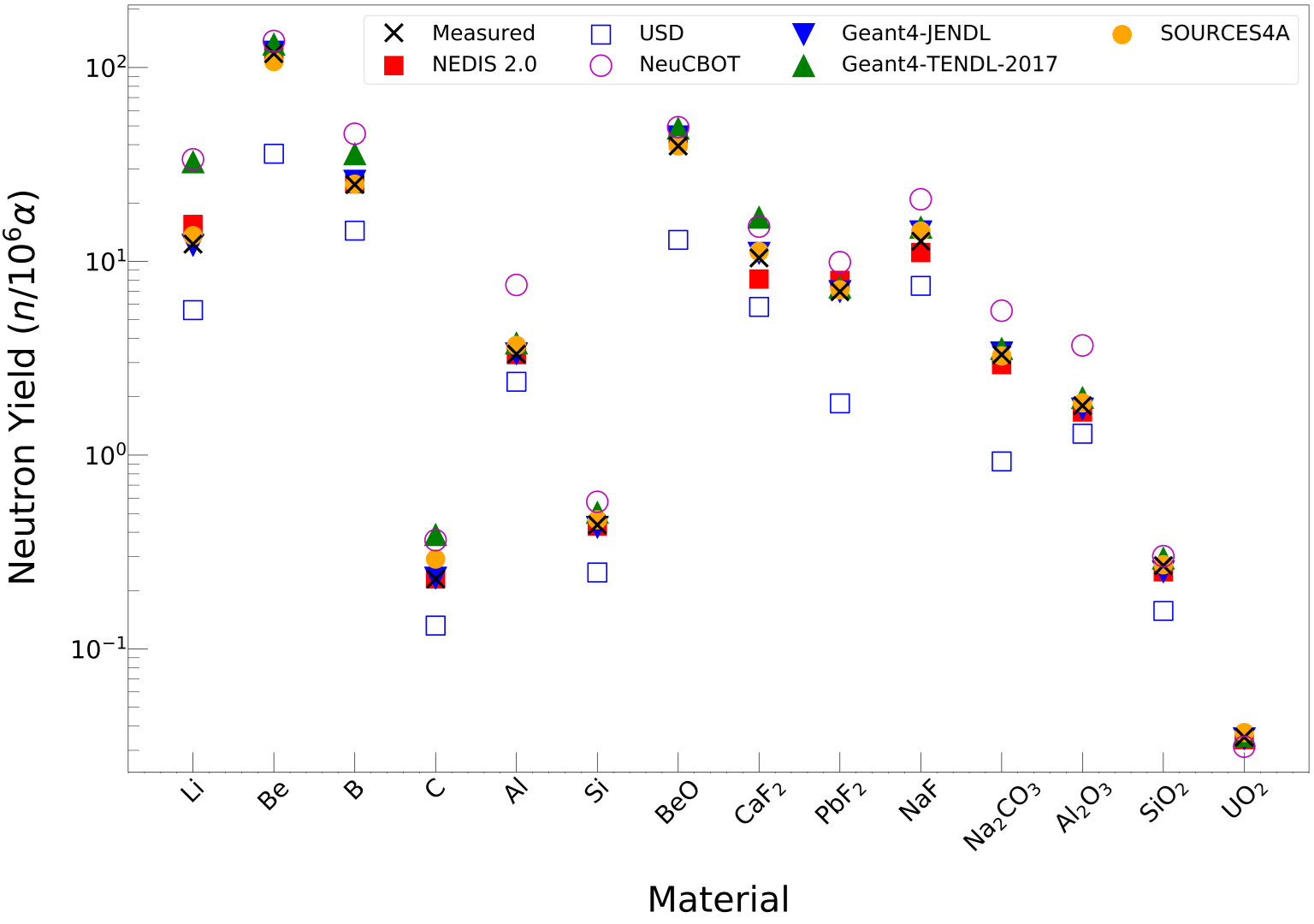}
\caption{Neutron yield $^{232}$Th decay chain in several materials as calculated by different codes in comparison with measurements. Neutron yield is given as the number of neutrons per $10^6$ alphas. See text for details.}
\label{fig:nyield-th232}
%\end{minipage}\hspace{0.5cm}
\end{figure}

Figure~\ref{fig:nyield-th232} shows the neutron yield for several materials from $^{232}$Th decay chain as calculated by different codes in comparison with experimental data. The neutron yield is given as the number of neutrons per $10^6$ alphas to be directly compared with other publications~\cite{fernandes,mendoza2019}. The decay chain is assumed to be in equilibrium. There are 6 alphas in the $^{232}$Th chain so to convert this to the total neutron production rate per gram per second per ppb of thorium the value from the figure needs to be multiplied by $6\times4.061\times10^{-6}/10^6$. SOURCES4A uses optimised cross-sections -- a combination of experimentally measured cross-sections and calculations either with TALYS1.9 or EMPIRE2.19/3.2.3. Data from the NEDIS-2.0 code were taken from Ref.~\cite{nedis}. USD data were obtained with the web-based toolkit developed at the University of South Dakota (USA)~\cite{usd} and reported in Ref.~\cite{fernandes}. NeuCBOT calculations have been reported in Ref.~\cite{neucbot}. Yields from GEANT4 simulations using two libraries JENDL~\cite{jendl} and TENDL-2017~\cite{talys} were calculated in Ref.~\cite{mendoza2019}. The `experimental data' were not directly measured as neutron yields from the whole decay chain but were evaluated from the measured neutron yields for different alpha energies as reported in Ref.~\cite{fernandes}. A good agreement between SOURCES4A and evaluated data is seen for most materials.

\begin{figure}[ht!]
%\begin{minipage}{8cm}
\includegraphics[width=16cm, height=8cm]{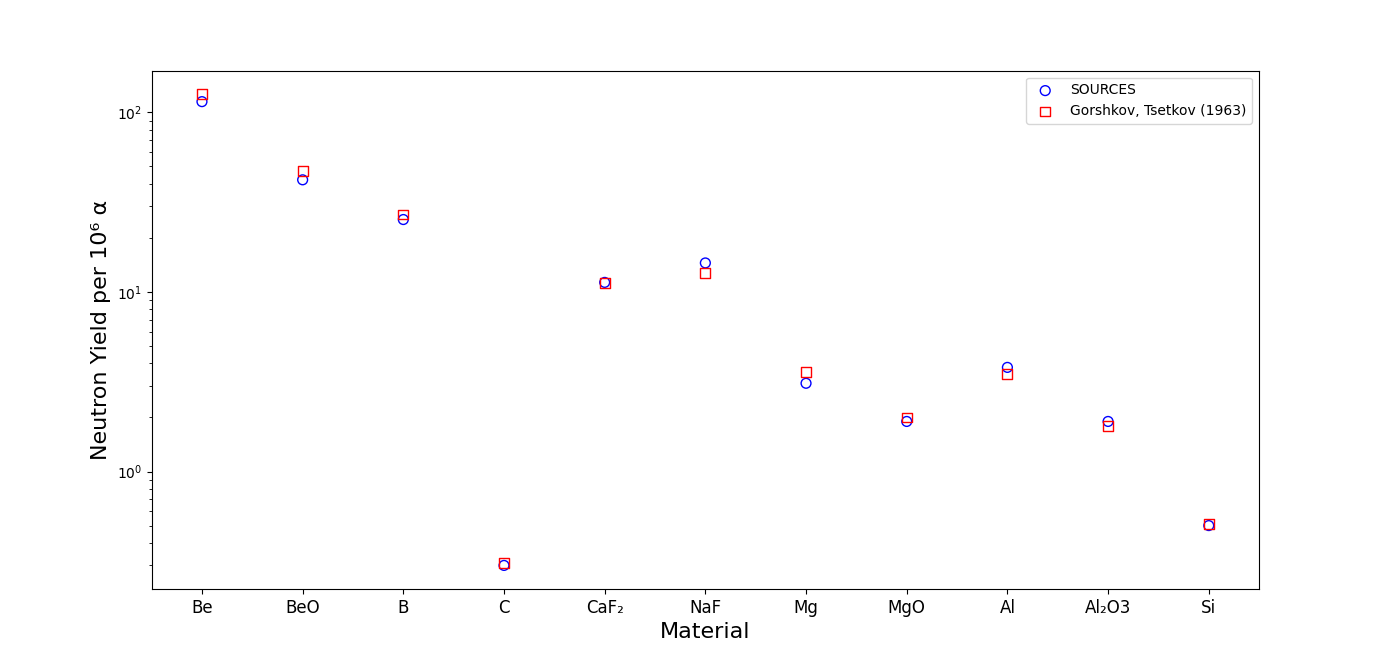}
\caption{Neutron yield $^{232}$Th decay chain in several materials as calculated by SOURCES4A in comparison with measurements~\cite{gorshkov}. Neutron yield is given as the number of neutrons per $10^6$ alphas.}
\label{fig:nyield-th232-gorshkov}
%\end{minipage}
\end{figure}

Ref.~\cite{gorshkov} reported direct measurement of neutron yields from uranium and thorium decay chains and the comparison of SOURCES4A output with these data for thorium chain is shown in Figure~\ref{fig:nyield-th232-gorshkov}. The calculations are again in very good agreement with measurements.

\section{Conclusions}
\label{sec:conclusions}

We have presented the `optimised' approach to the inputs to SOURCES4A when the cross-sections are taken from available data if available and consistency between different data sets is obvious, complemented with models from either TALYS or EMPIRE. With this input, we have calculated neutron yields as functions of alpha energy for a number of elements and materials and compared them to the measurements carried out with alpha particle beams. A good agreement is seen for most isotopes and materials tested. Calculated neutron energy spectra also show a reasonable agreement with the measurements. We have also presented a comparison of SOURCES4A calculations of neutron yields from radioactive decay chains with measurements and other codes. 

\begin{acknowledgments}

We acknowledge support from the UKRI-STFC and the University of Sheffield.

\end{acknowledgments}

%\nocite{*}
\bibliography{alphan-vk-lrt2022}% Produces the bibliography via BibTeX.

\end{document}